\documentclass[twocolumn,groupedaddress,
superscriptaddress,
showpacs,preprintnumbers,
amsmath,amssymb,
aps,prb,
longbibliography
]{revtex4-2}

\usepackage{newtxtext, newtxmath}

\usepackage{graphicx}
\usepackage{dcolumn}
\usepackage{bm}%
\usepackage[colorlinks=true,citecolor=blue]{hyperref}
\hypersetup{colorlinks=true,citecolor=blue,linkcolor=red,urlcolor=blue}

\usepackage{float}

\usepackage{color}
\definecolor{Green}{RGB}{0,204,102}
\definecolor{Purple}{RGB}{102,0,255}
\definecolor{Blue}{RGB}{51,153,255}
\definecolor{Red}{RGB}{151,010,010}

\begin{document}

\sloppy

\title{Excitons and Optical Response in Excitonic Insulator Candidate TiSe$_2$}


\newcommand*{\DIPC}[0]{{Donostia International Physics Center (DIPC), 20018 Donostia-San Sebasti\'an, Spain}}

\newcommand*{\IFS}[0]{{Centre for Advanced Laser Techniques, Institute of Physics, 10000 Zagreb, Croatia}}

\author{Dino Novko}
\email{dino.novko@gmail.com}
\affiliation{\IFS}
\affiliation{\DIPC}

\begin{abstract}
The origin of the charge density wave (CDW) phase in TiSe$_2$ is a highly debated topic, with lattice and excitonic correlations proposed as the main driving mechanisms. One of the proposed scenarios is the excitonic insulator (EI) mechanism, where soft electronic mode drives the phase transition. However, the existence of this purely electronic mode is controversial. Here, we perform fully ab-initio calculations of the electron excitation spectra in TiSe$_2$ with electron-hole excitonic effects included via Bethe-Salpeter equations. In the normal high-temperature phase the excitation spectra is dominated by the exciton mode at 1.6\,eV, while no well-defined soft electronic modes that could support the EI phase are present. In the CDW phase, the structural distortions induce a CDW band-gap opening between Ti-$d$ and Se-$p$ states, which supports the formation of the two low-energy excitonic modes in the optical spectrum at 0.4\,eV and 80\,meV. Close to the transition temperature $T_{\rm CDW}$, these two excitonic modes are softened and approach zero energy. These results suggest that the EI mechanism is not a main driving force in the formation of the CDW phase in TiSe$_2$, but there is a region in the phase diagram near $T_{\rm CDW}$ where EI fluctuations could be relevant.
\end{abstract}

\maketitle

Titanium diselenide (TiSe$_2$) is a widely discussed transition metal dichalcogenide (TMD) that hosts a charge density wave (CDW) phase below $T_{\rm CDW}=200$\,K, and is one of the earliest candidates for excitonic insulator (EI)\,\cite{disalvo76,kidd2002,cercellier07}, an intriguing state where soft collective electronic mode (i.e., plasmon or exciton) drives a phase transition\,\cite{jerome67}. In addition to purely electronic mechanism such as EI, electron-phonon coupling was often argued as an important contribution to the CDW phase transition\,\cite{rossnagel11,hughes1977,yoshida80,kidd2002,calandra11}.
In fact, recent ultrafast time-resolved measurements, in which electronic and lattice orders can be disentangled in time, have suggested that both EI and electron-phonon mechanisms cooperate in the formation of the CDW phase in TiSe$_2$\,\cite{rohwer2011collapse,vorobeva2011,hellmann2012time,otto2021mechanisms,cheng2022light,heinrich2023,fragkos2025}. While there are many experimental and theoretical studies clearly pointing to the structural instability and existence of the soft phonon mode that could drive the system to the CDW state, there are still many ambiguities regarding the corresponding soft electronic mode.

Below $T_{\rm CDW}$, Ti and Se atoms undergo periodic lattice distortions (PLDs) where unit cell is doubled in each crystal direction\,\cite{disalvo76}. In order for this to occur, there needs to be a soft, almost zero energy, phonon mode at the edge of the Brillouin zone (BZ), i.e., around $\mathbf{q}=\mathrm{M}$. This CDW-related phonon mode was observed both for $T<T_{\rm CDW}$ and $T>T_{\rm CDW}$ with x-ray (diffusive and inelastic) scattering measurements\,\cite{holt01,weber2011}, Raman and infrared spectroscopies\,\cite{holy1977,sugai1980,kogar2017a,velebit16}, as well as with time-resolved spectroscopic measurements as coherent phonons\,\cite{porer14,monney16,fragkos2025}. These observations are well supported by the model\,\cite{yoshida80,watanabe2015,kaneko2018} and density functional theory (DFT)\,\cite{calandra11,bianco2015electronic,hellgren2017critical,zhou2020anharmonicity,novko2022electron} calculations, showing that Ti-$d$ states at the edge of the BZ are strongly coupled to the Se-$p$ states at $\mathbf{k}=0$ via lattice motions, which in turn creates the temperature-dependent Kohn anomaly and hence the CDW instability.

On the other hand, the softening of the electronic mode at the edge of the BZ was observed by means of electron energy loss spectroscopy (EELS), and was used as an evidence for the EI scenario in TiSe$_2$\,\cite{kogar17}. However, in a recent EELS study this soft mode was ascribed to the CDW-related phonon, while it was shown that plasmons are strongly damped and non-existent away from the BZ center\,\cite{lin22}. Infrared spectroscopy study revealed another bulk plasmon around $\mathbf{q}=0$, not related to the electronic order, that is modulated across the CDW transition\,\cite{li2007semimetal}, and softening of the interband excitation peak at 0.4\,eV\,\cite{li2007semimetal,tyulnev2025}, which was also found in the resonant inelastic x-ray scattering (RIXS) measurements\,\cite{monney12}. The theoretical first-principles studies on these electronic excitations are similarly controversial. For instance, while the time-dependent DFT (TDDFT) study showed that a weak excitonic continuum is present at the edge of the BZ that is easily quenched by the PLDs and higher temperatures\,\cite{lian2019charge}, calculations based on model Bethe-Salpeter equations (BSE) report on a well-defined soft exciton that supports EI scenario\,\cite{pasquier18,tang2025}.
The rest of the theoretical studies on hybrid exciton-phonon mechanism discussed the impact of Coulomb interaction on the CDW phase diagram and $T_{\rm CDW}$, without reporting on the optical excitonic spectra and the existence of the CDW-related excitons\,\cite{wezel2010,monney10,monney2011,watanabe2015,monney15,chen18,kaneko2018,bok21,chen2023}. Therefore, it would be highly useful to explore the excitonic effects in TiSe$_2$ more deeply, which would not only broaden our understanding on the nature of the CDW in TiSe$_2$, but also provide some guidelines for other recently discovered EI candidates and their excitation spectra\,\cite{gao23,song23,chen23,gao24,kaneko25,golez25}.

\begin{figure*}[!t]
\begin{center}
\includegraphics[width=0.95\textwidth]{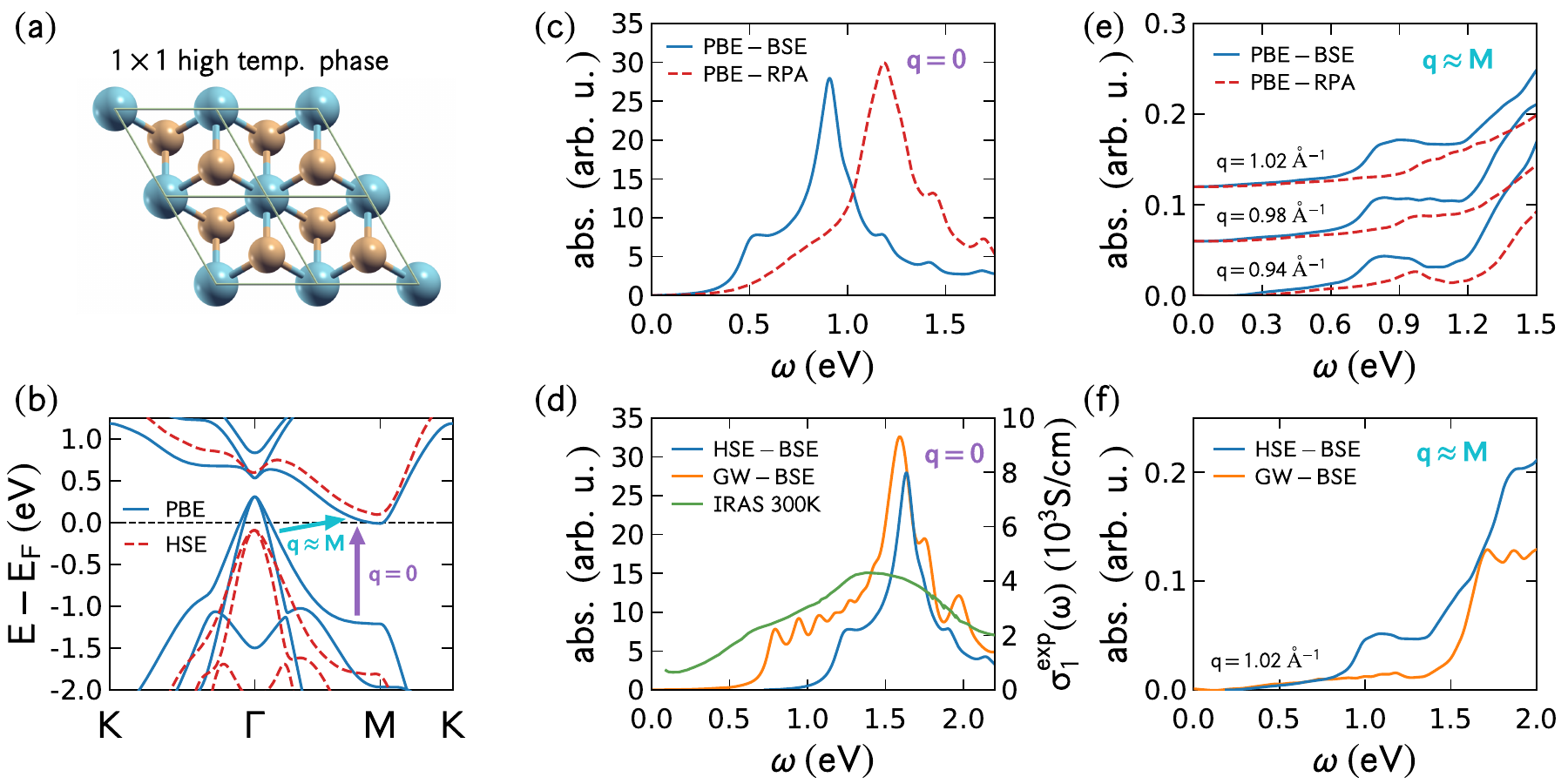}
\caption{(a) Crystal structure of TiSe$_2$ in the normal high-temperature phase. (b) Electronic band structure of TiSe$_2$ along high-symmetry points as calculated with DFT-PBE functional. The orange and blue arrows depict the vertical optical ($\mathbf{q}=0$) and finite-momentum low-energy ($\mathbf{q}=\mathrm{M}$) electron-hole transitions. (c) Optical absorption spectrum of TiSe$_2$ as calculated with BSE (that includes electron-hole interaction) and independent-particle RPA (no electron-hole interaction) approximations shown with blue and red dashed lines, respectively. These calculations were done on top of PBE electronic structure. (d) Optical absorption spectrum ($\mathbf{q}=0$) obtained with HSE-BSE and GW-BSE methods. The results are compared with optical absorption measurements of the high-temperature phase at $T=300$\,K\,\cite{li2007semimetal}.
(e) Absorption spectra obtained with BSE and RPA methods for three different momenta $\mathbf{q}$ close to the M point of the BZ. The topmost spectrum is calculated for $\mathbf{q}=\mathrm{M}$. 
 (f) Absorption spectra obtained with HSE-BSE and GW-BSE methods for $\mathbf{q}=\mathrm{M}$.
}
\label{fig1}
\end{center}
\end{figure*}

Here, we present the first-principles study on the excitation spectra across the phase diagram of TiSe$_2$  with excitonic (electron-hole interaction) effects included. The electron-hole interaction is incorporated into the excitation spectra by means of BSE calculations, while accurate electronic structure, such as CDW gap, is obtained with hybrid DFT functional and GW approximation\,\cite{cazzaniga2012,hellgren2017critical}. The theoretical zero momentum optical spectrum for high-temperature normal phase is dominated by the excitonic peak at around 1.6\,eV, in close agreement with the optical spectroscopy and EELS measurements\,\cite{buslaps1993,li2007semimetal,shi2019,tyulnev2025}, which originates from the vertical electron-hole excitations between Se-$p$ and Ti-$d$ states at the M point of the BZ. Contrary to some of the previous reports\,\cite{lian2019charge,pasquier18,tang2025}, finite momentum excitations do not show any well-defined low-energy or soft excitonic modes at the edge of the BZ, i.e., $\mathbf{q}\approx\mathrm{M}$, that would support the EI scenario of CDW formation in TiSe$_2$. In the low-temperature CDW phase, with $2\times2$ structure and PLDs due to electron-phonon coupling, we observe an opening of the CDW gap and formation of the low-energy exciton with a binding energy of around $300$\,meV, which corresponds to the 0.4\,eV interband peak observed previously in the experiments\,\cite{li2007semimetal,tyulnev2025}. The latter mode comes from the CDW gap excitations between the hybridized Ti-$d$ and Se-$p$ states close to the Fermi level. As the temperature increases towards $T_{\rm CDW}$, this low-energy exciton softens and quenches until the CDW gap is closed. Right below $T_{\rm CDW}$, another low-energy exciton is formed and it approaches zero energy even before the system goes from CDW to the normal state. These results suggest that the EI mechanism is not dominant, but probably a correcting factor, in the formation of the CDW phase. In fact, the EI state and EI fluctuations might be important near the $T_{\rm CDW}$ where the two observed excitonic modes approach zero energy.

Figures \ref{fig1}(a) and \ref{fig1}(b) show the crystal and band structures of TiSe$_2$ in the high-temperature normal phase. The DFT band structure obtained with PBE functional shows a semimetallic character, with hole-like Se-$p$ states at the $\Gamma$ and electron-like Ti-$d$ state at the M point of the BZ (for more computational details see Sec.\,S1 in Supplemental Material, SM\,\cite{SM}). When a more accurate DFT functional is used, such as hybrid HSE\,\cite{hse}, or when a GW approximation is applied, with a more accurate treatment of non-local electron interactions, electronic structure of TiSe$_2$ is renormalized with two key features (see also Fig.\,S1 in SM\,\cite{SM}). First, the indirect gap between Se-$p$ states at $\Gamma$ and Ti-$d$ states at M point is slightly increased. And second, the direct gap at the M point between these two bands is increased by around 0.7\,eV. These results are in line with previous calculations done with HSE and GW methods\,\cite{hellgren2017critical,cazzaniga2012,hellgren2021,acharya21} (see Fig.\,S1 for the comparison\,\cite{SM}) as well as with the meta-GGA approach\,\cite{yin24}.

The optical absorption spectrum (zero momentum excitation) up to 2\,eV as obtained with BSE on top of PBE electronic structure is dominated by an excitonic peak at around 0.9\,eV, which comes from the vertical electron-hole transitions at the M point of the BZ [see Fig.\,\ref{fig1}(c)], where Se-$p$ states exhibit a van Hove singularity and thus a large excitation density of states. By comparing the spectrum obtained with independent-particle RPA method with the BSE result, one finds that the binding energy of this exciton is around 300\,meV. When the beyond semi-local PBE band structure corrections are included (via HSE or GW approximations), the peak of this exciton is blueshifted to 1.6\,eV [see Fig.\,\ref{fig1}(d)], which is very close to the value of 1.37\,eV obtained in optical absorption spectroscopy\,\cite{li2007semimetal} and the value of 1.97\,eV obtained with EELS\,\cite{shi2019}. The excitonic feature observed in the experiments is, however, much more broader than the peak obtained here, which might be due to very strong electron scatterings of the Se-$p$ states at the M point. For instance, note that the states at M point at $-0.2$\,eV in photoemission experiments appear always very broad\,\cite{rohwer2011collapse,chen2015charge}. In addition, the exciton energy from optical absorption measurements is around 200\,meV below our theoretical estimate. This is reasonable since the inclusion of electron-phonon (or more precisely exciton-phonon) scattering would introduce both large broadening and redshift of the exciton energy\,\cite{novko2019,antonius2022}, providing a better agreement with the experiment. The value extracted from EELS is larger from both our theoretical result and optical absorption measurement, which is expected considering that EELS probes $-\mathrm{Im}\,1/\varepsilon(\omega)$, while the optical absorption is proportional to $\mathrm{Im}\,\varepsilon(\omega)$ [where $\varepsilon(\omega)$ is the dielectric function]. Note that this high-energy exciton is similar in nature to the exciton found in semimetallic graphene\,\cite{yang2009}, where strong electron-hole binding is formed also at the M point of the BZ where graphene $p$ states form a van Hove singularity.

\begin{figure*}[!t]
\begin{center}
\includegraphics[width=0.95\textwidth]{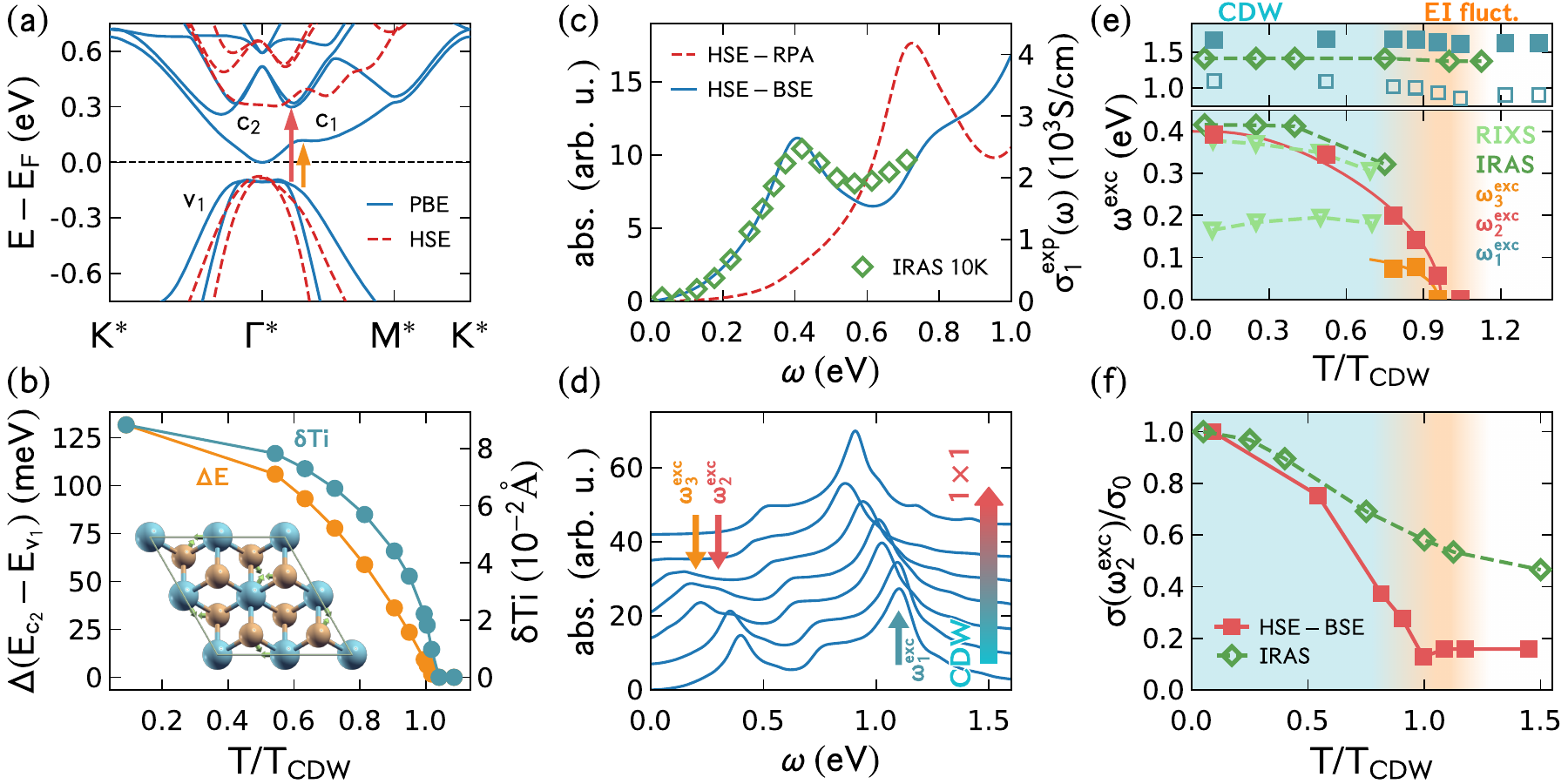}
\caption{(a) Electronic band structure of TiSe$_2$ in the CDW phase at 100\,K as obtained with PBE and HSE functionals. The CDW transition temperatures as obtained with PBE and HSE are 1100\,K and 2000\,K, respectively. The Se-$p$ states are denoted with v$_1$, while three backfolded Ti-$d$ states are labeled with c$_1$ and c$_2$ (note that the latter two are almost degenerate). Orange and red arrow show the electron-hole transitions that lead to the formation of low-energy excitons. (b) The energy of the CDW gap at the $\Gamma^{\ast}$ point and PLDs of Ti atoms as a function of temperature obtained with the PBE functional. Inset shows the $2\times 2$ CDW cell with directions of the PLDs. (c) Optical absorption spectra for the CDW phase at 100\,K as obtained with HSE-RPA and HSE-BSE methods. The experimental data is from Ref.\,\citenum{li2007semimetal}. (d) The HSE-BSE optical absorption spectra as a function of temperature across the CDW transition. Three excitonic peaks are labeled with arrows. (e) The energy peaks of the three excitonic modes as a function of temperature. The full and open blue squares are obtained with two different scissor corrections as explained in the text. The data are compared with optical absorption\,\cite{li2007semimetal} and RIXS\,\cite{monney12} results. The blue shaded area denotes the CDW phase dominated by the electron-phonon coupling, while the orange area approximately marks the range where EI fluctuations might be important. (f) The intensity of absorption at $\omega_{2}^{\rm exc}$ as a function of temperature and normalized to the value at the lowest temperature.}
\label{fig2}
\end{center}
\end{figure*}

The low-energy finite-momentum electron excitations close to $\mathbf{q}=\mathrm{M}$ are shown in Fig.\,\ref{fig1}(e) as obtained with PBE-BSE and PBE-RPA methods. Even though there are clear renormalizations of excitation spectra due to inclusion of electron-hole interactions (redshift of the absorption energy), there is no evidence of the well-defined excitonic mode. The same is true when the quasi-particle GW corrections are included or when the HSE hybrid functional is used, which accounts for a more accurate indirect gap between Ti-$d$ and Se-$p$ bands (see Fig.\,\ref{fig1}(f) and Fig.\,S2\,\cite{SM}). The EI mechanism in TiSe$_2$ assumes that Ti-$d$ electrons at the M point and Se-$p$ holes at $\Gamma$ are strongly bound and form excitons with finite momentum close to $\mathbf{q}=\mathrm{M}$. In the normal phase for $T\gtrsim T_{\rm CDW}$ this exciton should appear in excitation spectrum as a low-energy mode that softens with decreasing temperature. As mentioned, such mode was observed in the EELS measurements\,\cite{kogar17}, however, later it was argued that the soft mode is actually the CDW-related phonon mode\,\cite{lin22}. Recent model BSE calculations report the existence of this soft exciton close to the edge of the BZ\,\cite{pasquier18,tang2025}. However, here with a fully ab-initio BSE approach, we do not observe such temperature-dependent and well-defined exciton above $T_{\rm CDW}$ (see Figs.\,\ref{fig1}(e) and \ref{fig1}(f) as well as Fig.\,S3 in SM\,\cite{SM}), only a weak interband electron-hole continuum\,\cite{lian2019charge}. Given these observations it is unlikely that the EI scenario is the main mechanism of CDW formation in TiSe$_2$.

In Fig.\,\ref{fig2} we explore the optical response with excitonic effects below $T_{\rm CDW}$ when finite PLDs are present due to electron-phonon interaction. In these calculations we consider $2\times 2$ cell with distorted atoms along the eigenvectors of the CDW-related phonon mode and we relax the structure for each considered temperature (see SM for more details\,\cite{SM}). With PBE functional we get the CDW distorted structure until $T_{\rm CDW}^{\rm PBE}=1100$\,K, while with hybrid HSE functional we get $T_{\rm CDW}^{\rm HSE}=2000$\,K, which agrees well with the previous studies\,\cite{zhou2020anharmonicity,novko2022electron}. To obtain a more precise value for the transition temperature $T_{\rm CDW}$ one would need to account for the anharmonic effects\,\cite{zhou2020anharmonicity,chen2023}, which lowers the value to $\sim 400$\,K. Despite this discrepancy of $T_{\rm CDW}$, our results for the PLDs and electronic structure in the CDW phase, needed for accurate simulations of optical response, are in a very good agreement with the experimental observations, as shown below.

Figure \ref{fig2}(a) shows the electronic band structure around Fermi level in the CDW phase for the lowest considered temperature (maximal PLDs) as obtained with PBE and HSE functionals. Due to finite PLDs, the symmetry of the original unit cell is broken, and there is a significant hybridization (interaction) at the $\Gamma^{\ast}$ point between Se-$p$ and backfolded Ti-$d$ states. Here this strong coupling between these states is mediated through lattice motions, i.e., phonons. The corresponding CDW gap at the $\Gamma^{\ast}$ point between v$_1$ (Se-$p$) and c$_2$ (backfolded Ti-$p$) states is around 600\,meV and 700\,meV for PBE and HSE functionals, respectively. The CDW gap obtained with HSE functional is close to the one as obtained with GW approximation (See Fig.\,S4\,\cite{SM}), and agrees well with the previous reports\,\cite{zhou2020anharmonicity} (see also Fig.\,S5\,\cite{SM}). As shown in Fig.\,\ref{fig2}(b), as the temperature increases the value of PLDs and the CDW gap is decreasing, until it reaches zero for $T_{\rm CDW}$. Note that the PLDs obtained here are in an excellent agreement with the x-ray measurements\,\cite{fang2017}.

The low-energy optical absorption data for the CDW phase (lowest $T$ and largest PLDs) is depicted in Fig.\,\ref{fig2}(c) as obtained with BSE and RPA methods. Here we use a scissor method to adjust the value of the CDW gap to the HSE value (see Fig.\,S5\,\cite{SM}). The low-energy spectra show an excitonic peak at 0.4\,eV, with strong binding energy of about 300\,meV. This value is very close to the exciton binding energies of semiconducting TMDs, which are regarded as materials with promising excitonic properties\,\cite{manzeli17,wang2018}. Similar results are also obtained with the GW-BSE approach (see Fig.\,S6\,\cite{SM}). Our results are in an excellent agreement with the low-temperature infrared absorption spectroscopy\,\cite{li2007semimetal,tyulnev2025}. This excitonic mode comes from the coupling between hybridized v$_1$ and c$_2$ states, slightly away from the $\Gamma^{\ast}$ point, as pointed with the red arrow in Fig.\,\ref{fig2}(a).

In Fig.\,\ref{fig2}(d) we present the optical absorption spectra across the CDW phase transition, shown in a broader energy range. As the temperature increases towards $T_{\rm CDW}$, the CDW gap is closing, and hence the low-energy exciton is redshifted and quenched. Besides the excitonic peak ($\omega_{1}^{\rm exc}$) at around 1\,eV that is also present in the $1\times 1$ high-temperature phase [see Figs.\,\ref{fig1}(c) and \ref{fig1}(d)], and the 0.4\,eV exciton that comes from the opening of the CDW gap ($\omega_{2}^{\rm exc}$), we observe another exciton ($\omega_{3}^{\rm exc}$) that is formed close to $T_{\rm CDW}$. This mode comes from the coupling between v$_1$ and c$_1$ states [see the orange arrow in Fig.\,\ref{fig2}(a)], and approaches zero energy even before the CDW transitions to the normal, undistorted phase. This hints that there could be an EI phase or EI fluctuations close to the transition temperature $T_{\rm CDW}$. Figure \ref{fig2}(e) shows the energies of these three excitons as a function of temperature, and compared with the corresponding modes as observed in optical absorption\,\cite{li2007semimetal} and RIXS\,\cite{monney12} studies. First of all, note that the states that form exciton $\omega_{1}^{\rm exc}$ renormalize very differently than the low-energy states and the CDW gap (compare Figs.\,S7 and S5\,\cite{SM}). In other words, the Se-$p$ states at the M point of the BZ and Ti-$d$ c$_1$ states that form the $\omega_{1}^{\rm exc}$ exciton are not affected by the formation of the CDW. Therefore, a different scissor method needs to be used in order to describe the temperature dependence of $\omega_{1}^{\rm exc}$. When this is properly taken into account, we get a temperature-independent exciton at $\omega_{1}^{\rm exc}=1.6$\,eV, in a close agreement with the experiments\,\cite{li2007semimetal,tyulnev2025}. The temperature dependence of excitonic mode $\omega_{2}^{\rm exc}$ follows the behavior of the CDW gap [see Fig.\,\ref{fig2}(b)], typical for the second-order phase transition. The energy positions of this mode are in very good agreement with the optical absorption and RIXS studies. The optical absorption measurements do not show a second low-energy mode $\omega_{3}^{\rm exc}$, while RIXS reports a second mode between 0.1\,eV and 0.2\,eV that might be related to this exciton. Furthermore, we show the intensity of absorption at the peak of the $\omega_{2}^{\rm exc}$ exciton compared to the experiments\,\cite{li2007semimetal,tyulnev2025}. While in our calculations the low-energy spectral weight is completely diminished with a closing of the CDW gap for $T\geq T_{\rm CDW}$, the intensity in the experiments is still significant far above $T_{\rm CDW}$. The missing intensity might come from the increase of the electron-phonon scattering channels for $T\geq T_{\rm CDW}$\,\cite{torbatian2024,velebit16} that could lead to indirect phonon-assisted absorption\,\cite{giustino17}. Another likely possibility is that the finite intensity for $T\geq T_{\rm CDW}$ points to the existence of the CDW fluctuations, which are documented to persists well above $T_{\rm CDW}$\,\cite{holt01,kidd2002,cercellier07,monney16,cheng2022light,fragkos2025}. Within this CDW fluctuating phase, a finite fluctuating PLDs and CDW gap are present due to soft phonon mode and strong dynamical electron-phonon scattering\,\cite{fragkos2025}, and therefore, it is possible that the low-energy excitons $\omega_{2}^{\rm exc}$ and $\omega_{3}^{\rm exc}$ are extended beyond $T_{\rm CDW}$. In other words, this part of the phase diagram, where the experiments are reporting finite absorption intensity, could be a region where soft CDW phonon mode and soft excitons are hybridized and influence each other.

In conclusion, we have studied the electron excitation spectra across the CDW transition in TiSe$_2$, with electron-hole interactions included. The results for the optical absorption in the high-temperature normal phase show a strong exciton at 1.6\,eV, in a close agreement with the experiments. At finite momenta, where soft electronic mode is expected, supporting the excitonic insulator scenario, we found no evidence of the well-defined low-energy mode. Below the CDW transition temperature, we have revealed two additional low-energy excitons, which are formed from the hybridized Ti-$d$ and Se-$p$ states at the center of the Brillouin zone. These two modes soften towards the CDW transition, and are quenched in the normal phase. The softening and quenching of excitons is related to the closing of the CDW electronic gap and reduction of the phonon-induced lattice distortions. Even though these results suggest that excitonic insulator correlations do not play a dominant role in the formation of the CDW in TiSe$_2$, the existence of the soft excitonic modes close to the CDW transition temperature reveal a small region of the phase diagram where excitonic-insulator fluctuations might be active. These results might be useful for understanding the nature of the phase transitions in other excitonic insulator candidates, where also the respective roles of lattice and excitonic correlations are similarly ambiguous\,\cite{chen23,gao23,song23,gao24,samaneh2021,sabatino2023,kaneko25}.
Beyond the fundamental importance, these low-energy excitonic modes with strong binding energy might be utilized for enhancing the light-matter interaction, as it is done, for instance, in the CDW-bearing TaSe$_2$\,\cite{song2021plasmons} and kagome metal CsV$_3$Sb$_5$\,\cite{meng2025} where strong exciton-plasmon coupling is observed across the CDW transition\,\cite{torbatian2024,zhou2024}.

\begin{acknowledgments}
D.N. acknowledges financial support from the Croatian Science Foundation (Grant no. IP-2025-02-5926), from the European Regional Development Fund for the ``Center of Excellence for Advanced Materials and Sensing Devices'' (Grant No. KK.01.1.1.01.0001), and from the project "Podizanje znanstvene izvrsnosti Centra za napredne laserske tehnike (CALTboost)" financed by the European Union through the National Recovery and Resilience Plan 2021-2026 (NRPP). Computational resources were provided by the DIPC computing center.
\end{acknowledgments}

\bibliography{tise2}

\end{document}